%% file: main.tex
\pdfoutput=1

\documentclass[12pt,a4paper]{article}
\newcommand{\TwoColumns}{False}

\usepackage{lineno}  % for line numbering during review
\usepackage{graphicx}  % to include figures (can also use other packages)
\usepackage{xspace} % To avoid problems with missing or double spaces after
\usepackage{color}
\usepackage{colortbl}
\usepackage{amsmath} % Adds a large collection of math symbols
\usepackage{ifthen} % for conditional statements

\usepackage{multirow}

\newboolean{pdflatex}
\setboolean{pdflatex}{true} % use this if using non-eps figures
\newboolean{articletitles}
\setboolean{articletitles}{true} % False removes titles in references

\newboolean{uprightparticles}
\setboolean{uprightparticles}{false} %Set to true to get roman particle symbols

\usepackage{amssymb}
\usepackage{amsfonts}
\usepackage{upgreek} % Adds in support for greek letters in roman typeset

\usepackage{hyperref}    % Hyperlinks in references
\usepackage[all]{hypcap} % Internal hyperlinks to floats.

\input{lhcb-symbols-def} % Add in the predefined LHCb symbols

\usepackage{mciteplus}

\begin{document}

%%%%%%%%%%%%%%%%%%%%%%%%%
%%%%% Title     %%%%%%%%%
%%%%%%%%%%%%%%%%%%%%%%%%%

\renewcommand{\thefootnote}{\fnsymbol{footnote}}
\setcounter{footnote}{1}

\input{title-LHCb-PAPER}

\renewcommand{\thefootnote}{\arabic{footnote}}
\setcounter{footnote}{0}

%%%%%%%%%%%%%%%%%%%%%%%%%%%%%%%%
%%%%%  Table of Content   %%%%%%
%%%%%%%%%%%%%%%%%%%%%%%%%%%%%%%%
%%%% Uncomment next 2 lines if desired
%\tableofcontents
%\cleardoublepage

%%%%%%%%%%%%%%%%%%%%%%%%%
%%%%% Main text %%%%%%%%%
%%%%%%%%%%%%%%%%%%%%%%%%%

\pagestyle{plain} % restore page numbers for the main text
\setcounter{page}{1}
\pagenumbering{arabic}

\input{introduction}

\input{scale}

\input{alt-selection}

\input{fits}

\input{systematics}

\input{conclusion}
\input{acknowledgements}

% \clearpage
\bibliographystyle{LHCb}
\bibliography{main}

\end{document}

%% file: lhcb-symbols-def.tex
\def\lhcb {\mbox{LHCb}\xspace}
\def\ux85 {\mbox{UX85}\xspace}

\def\cdf    {\mbox{CDF}\xspace}
\def\dzero  {\mbox{D0}\xspace}

\ifthenelse{\boolean{uprightparticles}}%
{
 
 \def\Pgamma      {\ensuremath{\upgamma}\xspace}

 \def\Peta        {\ensuremath{\upeta}\xspace}

 \def\Pmu         {\ensuremath{\upmu}\xspace}                 
 \def\Pnu         {\ensuremath{\upnu}\xspace}                 
                  
 \def\Ppi         {\ensuremath{\uppi}\xspace}

 \def\Ptau        {\ensuremath{\uptau}\xspace}

 \def\Pchi        {\ensuremath{\upchi}\xspace}                 
 \def\Ppsi        {\ensuremath{\uppsi}\xspace}

 \def\PDelta      {\ensuremath{\Delta}\xspace}                 
 \def\PXi      {\ensuremath{\Xi}\xspace}                 
 \def\PLambda      {\ensuremath{\Lambda}\xspace}                 
 \def\PSigma      {\ensuremath{\Sigma}\xspace}                 
 \def\POmega      {\ensuremath{\Omega}\xspace}                 
 \def\PUpsilon      {\ensuremath{\Upsilon}\xspace}                 
 
 \def\PB      {\ensuremath{\mathrm{B}}\xspace}                 
                  
 \def\PD      {\ensuremath{\mathrm{D}}\xspace}

 \def\PH      {\ensuremath{\mathrm{H}}\xspace}                 
                  
 \def\PJ      {\ensuremath{\mathrm{J}}\xspace}                 
 \def\PK      {\ensuremath{\mathrm{K}}\xspace}

 \def\PW      {\ensuremath{\mathrm{W}}\xspace}

 \def\PZ      {\ensuremath{\mathrm{Z}}\xspace}                 
                  
 \def\Pb      {\ensuremath{\mathrm{b}}\xspace}                 
 \def\Pc      {\ensuremath{\mathrm{c}}\xspace}                 
 \def\Pd      {\ensuremath{\mathrm{d}}\xspace}                 
 \def\Pe      {\ensuremath{\mathrm{e}}\xspace}

 \def\Ph      {\ensuremath{\mathrm{h}}\xspace}                 
 \def\Pi      {\ensuremath{\mathrm{i}}\xspace}

 \def\Pn      {\ensuremath{\mathrm{n}}\xspace}                 
                  
 \def\Pp      {\ensuremath{\mathrm{p}}\xspace}                 
 \def\Pq      {\ensuremath{\mathrm{q}}\xspace}                 
                  
 \def\Ps      {\ensuremath{\mathrm{s}}\xspace}                 
 \def\Pt      {\ensuremath{\mathrm{t}}\xspace}                 
 \def\Pu      {\ensuremath{\mathrm{u}}\xspace}

}
{
 
 \def\Pgamma      {\ensuremath{\gamma}\xspace}

 \def\Peta        {\ensuremath{\eta}\xspace}

 \def\Pmu         {\ensuremath{\mu}\xspace}                 
 \def\Pnu         {\ensuremath{\nu}\xspace}                 
                  
 \def\Ppi         {\ensuremath{\pi}\xspace}

 \def\Ptau        {\ensuremath{\tau}\xspace}

 \def\Pchi        {\ensuremath{\chi}\xspace}                 
 \def\Ppsi        {\ensuremath{\psi}\xspace}                 
                  
 \mathchardef\PDelta="7101
 \mathchardef\PXi="7104
 \mathchardef\PLambda="7103
 \mathchardef\PSigma="7106
 \mathchardef\POmega="710A
 \mathchardef\PUpsilon="7107
                  
 \def\PB      {\ensuremath{B}\xspace}                 
                  
 \def\PD      {\ensuremath{D}\xspace}

 \def\PH      {\ensuremath{H}\xspace}                 
                  
 \def\PJ      {\ensuremath{J}\xspace}                 
 \def\PK      {\ensuremath{K}\xspace}

 \def\PW      {\ensuremath{W}\xspace}

 \def\PZ      {\ensuremath{Z}\xspace}                 
                  
 \def\Pb      {\ensuremath{b}\xspace}                 
 \def\Pc      {\ensuremath{c}\xspace}                 
 \def\Pd      {\ensuremath{d}\xspace}                 
 \def\Pe      {\ensuremath{e}\xspace}

 \def\Ph      {\ensuremath{h}\xspace}                 
 \def\Pi      {\ensuremath{i}\xspace}

 \def\Pn      {\ensuremath{n}\xspace}                 
                  
 \def\Pp      {\ensuremath{p}\xspace}                 
 \def\Pq      {\ensuremath{q}\xspace}                 
                  
 \def\Ps      {\ensuremath{s}\xspace}                 
 \def\Pt      {\ensuremath{t}\xspace}                 
 \def\Pu      {\ensuremath{u}\xspace}

}

\def\en         {\ensuremath{\Pe^-}\xspace}   % electron negative (\em is taken)

\def\epem       {\ensuremath{\Pe^+\Pe^-}\xspace}

\def\mup        {\ensuremath{\Pmu^+}\xspace}
\def\mun        {\ensuremath{\Pmu^-}\xspace} % muon negative (\mum is taken)

\def\neu        {\ensuremath{\Pnu}\xspace}
\def\neub       {\ensuremath{\overline{\Pnu}}\xspace}

\def\neue       {\ensuremath{\neu_e}\xspace}
\def\neueb      {\ensuremath{\neub_e}\xspace}

\def\neum       {\ensuremath{\neu_\mu}\xspace}
\def\neumb      {\ensuremath{\neub_\mu}\xspace}

\def\neut       {\ensuremath{\neu_\tau}\xspace}
\def\neutb      {\ensuremath{\neub_\tau}\xspace}

\def\neul       {\ensuremath{\neu_\ell}\xspace}
\def\neulb      {\ensuremath{\neub_\ell}\xspace}

\def\g      {\ensuremath{\Pgamma}\xspace}

\def\dquark    {\ensuremath{\Pd}\xspace}

\def\squark    {\ensuremath{\Ps}\xspace}

\def\cquark    {\ensuremath{\Pc}\xspace}

\def\bquark    {\ensuremath{\Pb}\xspace}

\def\pion  {\ensuremath{\Ppi}\xspace}

\def\pip   {\ensuremath{\pion^+}\xspace}
\def\pim   {\ensuremath{\pion^-}\xspace}

\def\kaon  {\ensuremath{\PK}\xspace}
  \def\Kbar  {\kern 0.2em\overline{\kern -0.2em \PK}{}\xspace}

\def\Kz    {\ensuremath{\kaon^0}\xspace}
\def\Kzb   {\ensuremath{\Kbar^0}\xspace}
\def\KzKzb {\ensuremath{\Kz \kern -0.16em \Kzb}\xspace}
\def\Kp    {\ensuremath{\kaon^+}\xspace}
\def\Km    {\ensuremath{\kaon^-}\xspace}

\def\KpKm  {\ensuremath{\Kp \kern -0.16em \Km}\xspace}
\def\KS    {\ensuremath{\kaon^0_{\rm\scriptscriptstyle S}}\xspace} 
 
\def\Kstarz  {\ensuremath{\kaon^{*0}}\xspace}
\def\Kstarzb {\ensuremath{\Kbar^{*0}}\xspace}

  \def\Dbar    {\kern 0.2em\overline{\kern -0.2em \PD}{}\xspace}
\def\D       {\ensuremath{\PD}\xspace}

\def\Dz      {\ensuremath{\D^0}\xspace}
\def\Dzb     {\ensuremath{\Dbar^0}\xspace}
\def\DzDzb   {\ensuremath{\Dz {\kern -0.16em \Dzb}}\xspace}
\def\Dp      {\ensuremath{\D^+}\xspace}
\def\Dm      {\ensuremath{\D^-}\xspace}

\def\DpDm    {\ensuremath{\Dp {\kern -0.16em \Dm}}\xspace}

\def\B       {\ensuremath{\PB}\xspace}
  \def\Bbar    {\kern 0.18em\overline{\kern -0.18em \PB}{}\xspace}

\def\Bu      {\ensuremath{\B^+}\xspace}
\def\Bub     {\ensuremath{\B^-}\xspace}
\def\Bp      {\ensuremath{\Bu}\xspace}

\def\Bd      {\ensuremath{\B^0}\xspace}
\def\Bs      {\ensuremath{\B^0_\squark}\xspace}

\def\Bdb     {\ensuremath{\Bbar^0}\xspace}
\def\Bc      {\ensuremath{\B_\cquark^+}\xspace}

\def\jpsi     {\ensuremath{{\PJ\mskip -3mu/\mskip -2mu\Ppsi\mskip 2mu}}\xspace}

  \def\Y#1S{\ensuremath{\PUpsilon{(#1S)}}\xspace}% no space before {...}!

\def\proton      {\ensuremath{\Pp}\xspace}

\def\neutron     {\ensuremath{\Pn}\xspace}

\def\Deltares {\ensuremath{\PDelta}\xspace}

\def\Xires {\ensuremath{\PXi}\xspace}

\def\L {\ensuremath{\PLambda}\xspace}
\def\Lbar {\ensuremath{\kern 0.1em\overline{\kern -0.1em\PLambda}}\xspace}
\def\Lambdares {\ensuremath{\PLambda}\xspace}

\def\Sigmares {\ensuremath{\PSigma}\xspace}

\def\Omegares {\ensuremath{\POmega}\xspace}

\def\Lb      {\ensuremath{\L^0_\bquark}\xspace}

\def\myL {\ensuremath{\PLambda}\xspace}
\def\myLb {\ensuremath{\PLambda^0_\bquark}\xspace}
\def\myX {\ensuremath{\PXi^-}\xspace}

\def\myXb {\ensuremath{\PXi^-_\bquark}\xspace}
\def\myO {\ensuremath{\POmega^-}\xspace}
\def\myOb {\ensuremath{\POmega^-_\bquark}\xspace}

\def\myXcz {\ensuremath{\PXi^0_\cquark}\xspace}

\newcommand{\decay}[2]{\ensuremath{#1\!\to #2}\xspace}         % {\Pa}{\Pb \Pc}

\def\to                 {\ensuremath{\rightarrow}\xspace}

\newcommand{\phid}{\ensuremath{\phi_{\dquark}}\xspace}

\newcommand{\phis}{\ensuremath{\phi_{\squark}}\xspace}

\newcommand{\etag}{{\ensuremath{\varepsilon_{\rm tag}}}\xspace}

\def\AT#1     {\ensuremath{A_{\mathrm{T}}^{#1}}\xspace}           % 2

\def\C#1      {\ensuremath{\mathcal{C}_{#1}}\xspace}                       % 9
\def\Cp#1     {\ensuremath{\mathcal{C}_{#1}^{'}}\xspace}                    % 7
\def\Ceff#1   {\ensuremath{\mathcal{C}_{#1}^{\mathrm{(eff)}}}\xspace}        % 9  
\def\Cpeff#1  {\ensuremath{\mathcal{C}_{#1}^{'\mathrm{(eff)}}}\xspace}       % 7
\def\Ope#1    {\ensuremath{\mathcal{O}_{#1}}\xspace}                       % 2
\def\Opep#1   {\ensuremath{\mathcal{O}_{#1}^{'}}\xspace}                    % 7

\newcommand{\unit}[1]{\ensuremath{\rm\,#1}\xspace}          % {kg}

\newcommand{\tev}{\ensuremath{\mathrm{\,Te\kern -0.1em V}}\xspace}
\newcommand{\gev}{\ensuremath{\mathrm{\,Ge\kern -0.1em V}}\xspace}
\newcommand{\mev}{\ensuremath{\mathrm{\,Me\kern -0.1em V}}\xspace}
\newcommand{\kev}{\ensuremath{\mathrm{\,ke\kern -0.1em V}}\xspace}
\newcommand{\ev}{\ensuremath{\mathrm{\,e\kern -0.1em V}}\xspace}
\newcommand{\gevc}{\ensuremath{{\mathrm{\,Ge\kern -0.1em V\!/}c}}\xspace}
\newcommand{\mevc}{\ensuremath{{\mathrm{\,Me\kern -0.1em V\!/}c}}\xspace}
\newcommand{\gevcc}{\ensuremath{{\mathrm{\,Ge\kern -0.1em V\!/}c^2}}\xspace}
\newcommand{\gevgevcccc}{\ensuremath{{\mathrm{\,Ge\kern -0.1em V^2\!/}c^4}}\xspace}
\newcommand{\mevcc}{\ensuremath{{\mathrm{\,Me\kern -0.1em V\!/}c^2}}\xspace}

\def\mum  {\ensuremath{\,\upmu\rm m}\xspace}

\def\gsim{{~\raise.15em\hbox{$>$}\kern-.85em
          \lower.35em\hbox{$\sim$}~}\xspace}
\def\lsim{{~\raise.15em\hbox{$<$}\kern-.85em
          \lower.35em\hbox{$\sim$}~}\xspace}

\def\tell1  {TELL1\xspace}
\def\ukl1   {UKL1\xspace}

%% file: title-LHCb-PAPER.tex
 %%%%%%%%%%%%%%%%%%%%%%%%%
 %%%%%  TITLE PAGE  %%%%%%
 %%%%%%%%%%%%%%%%%%%%%%%%%
 \begin{titlepage}
 \pagenumbering{roman}
 
% Header ---------------------------------------------------
 \vspace*{-1.5cm}
 \centerline{\large EUROPEAN ORGANIZATION FOR NUCLEAR RESEARCH (CERN)}
 \vspace*{1.5cm}
 \hspace*{-0.5cm}
 \begin{tabular*}{\linewidth}{lc@{\extracolsep{\fill}}r}
 \ifthenelse{\boolean{pdflatex}}% Logo format choice
 {\vspace*{-2.7cm}\mbox{\!\!\!\includegraphics[width=.14\textwidth]{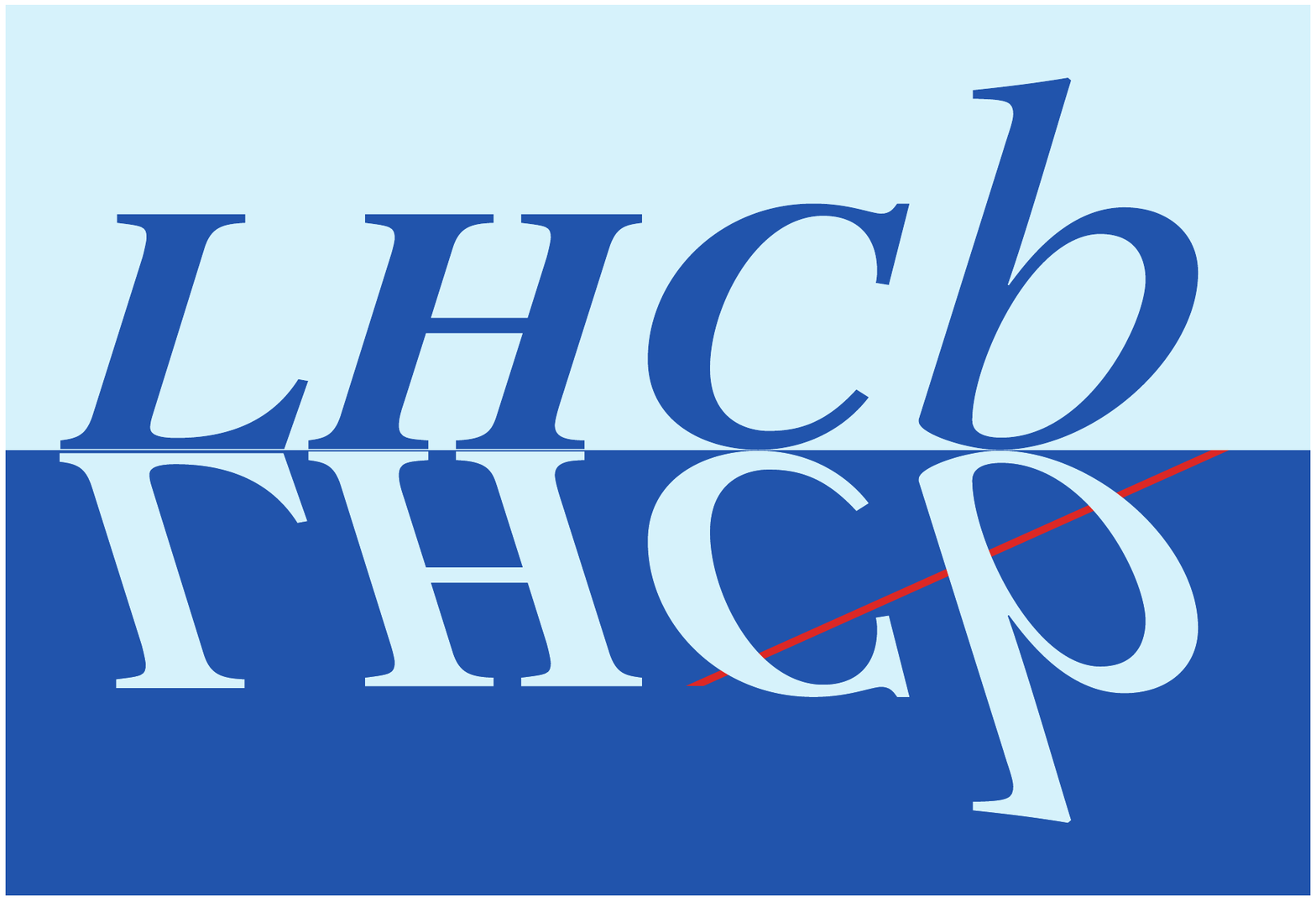}} & &}%
 {\vspace*{-1.2cm}\mbox{\!\!\!\includegraphics[width=.12\textwidth]{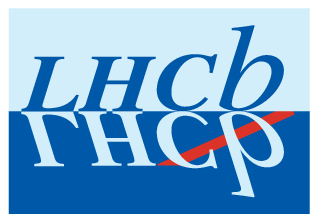}} & &}%
 \\
 & & CERN-PH-EP-2013-013 \\  % ID 
 & & LHCb-PAPER-2012-048 \\  % ID 
 & & February 5, 2013 \\ % Date
  & & \\
 % not in paper \hline
 \end{tabular*}
 
\vspace*{2.0cm}
 
% Title --------------------------------------------------
 {\bf\boldmath\huge
 \begin{center}
   Measurement of the \myLb, \myXb and \myOb baryon masses
 \end{center}
 }
 
\vspace*{1.75cm}
 
% Authors -------------------------------------------------
 \begin{center}
 The LHCb collaboration\footnote{Authors are listed on the following pages.}
 \end{center}
 
\vspace{\fill}
 
% Abstract -----------------------------------------------
 \begin{abstract}
   \noindent
Bottom baryons decaying to a \jpsi meson and a hyperon are reconstructed
using $\rm 1.0~fb^{-1}$ of data collected in 2011 with the LHCb detector.
Significant $\myLb\to\jpsi\myL$,
$\myXb\to\jpsi\myX$ and $\myOb\to\jpsi\myO$ signals
are observed and the corresponding masses are 
measured to be 
\begin{center}
\begin{tabular}{l @{$~=~$}l@{$\,\,\pm\,$}l@{\,(stat) $\pm\,$}l@{\,(syst) MeV/$c^2$}l}
$M(\myLb)$  & 5619.53 & 0.13 & 0.45 & , \\
$M(\myXb)$  & 5795.8  & 0.9  & 0.4  & , \\
$M(\myOb)$  & 6046.0  & 2.2  & 0.5  & , \\
\end{tabular}
\end{center}
while the differences with respect to the \myLb mass are
\begin{center}
\begin{tabular}{l @{$~=~$}l@{$\,\,\pm\,$}l@{\,(stat) $\pm\,$}l@{\,(syst) MeV/$c^2$}l}
$M(\myXb) -M(\myLb)$ & 176.2  & 0.9  & 0.1  & , \\
$M(\myOb) -M(\myLb)$ & 426.4  & 2.2  & 0.4  & . \\
\end{tabular}
\end{center}
These are the most precise mass measurements of the \myLb, \myXb and \myOb baryons to date.
Averaging the above \myLb mass measurement with that published by \lhcb\ using $\rm 35~pb^{-1}$ of data collected in 2010 yields 
$M(\myLb) = \rm 5619.44 \pm 0.13\,(stat) \pm 0.38\,(syst)$\,\mevcc.
 \end{abstract}
 
\vspace*{1cm}
 
\begin{center}
   Submitted to Physical Review Letters
\end{center}
 
\vspace{\fill}
 
{\footnotesize 
\centerline{\copyright~CERN on behalf of the \lhcb collaboration, license \href{http://creativecommons.org/licenses/by/3.0/}{CC-BY-3.0}.}}
 \vspace*{2mm}
 
\end{titlepage}

 %%%%%%%%%%%%%%%%%%%%%%%%%%%%%%%%
 %%%%%  EOD OF TITLE PAGE  %%%%%%
 %%%%%%%%%%%%%%%%%%%%%%%%%%%%%%%%
 
%  empty page follows the title page ----
 \newpage
 \setcounter{page}{2}
 \mbox{~}
 \newpage
 
% Author List ----------------------------
 \input{LHCb_authorlist.tex}

\cleardoublepage

%% file: LHCb_authorlist.tex
%%%%%%%%%%%%%%%%%%%%%%%%%%%%%%%%%%%%%%%%%%
\centerline{\large\bf LHCb collaboration}
\begin{flushleft}
\small
R.~Aaij$^{38}$, 
C.~Abellan~Beteta$^{33,n}$, 
A.~Adametz$^{11}$, 
B.~Adeva$^{34}$, 
M.~Adinolfi$^{43}$, 
C.~Adrover$^{6}$, 
A.~Affolder$^{49}$, 
Z.~Ajaltouni$^{5}$, 
J.~Albrecht$^{9}$, 
F.~Alessio$^{35}$
% ...

{\footnotesize \it
% ...
$ ^{5}$Clermont Universit\'{e}, Universit\'{e} Blaise Pascal, CNRS/IN2P3, LPC, Clermont-Ferrand, France\\
$ ^{6}$CPPM, Aix-Marseille Universit\'{e}, CNRS/IN2P3, Marseille, France\\
% ...
$ ^{9}$Fakult\"{a}t Physik, Technische Universit\"{a}t Dortmund, Dortmund, Germany\\
% ...
$ ^{11}$Physikalisches Institut, Ruprecht-Karls-Universit\"{a}t Heidelberg, Heidelberg, Germany\\
% ...
$ ^{33}$Universitat de Barcelona, Barcelona, Spain\\
$ ^{34}$Universidad de Santiago de Compostela, Santiago de Compostela, Spain\\
$ ^{35}$European Organization for Nuclear Research (CERN), Geneva, Switzerland\\
% ...
$ ^{38}$Nikhef National Institute for Subatomic Physics, Amsterdam, The Netherlands\\
% ...
$ ^{43}$H.H. Wills Physics Laboratory, University of Bristol, Bristol, United Kingdom\\
% ...
$ ^{49}$Oliver Lodge Laboratory, University of Liverpool, Liverpool, United Kingdom\\
% ...
$ ^{n}$LIFAELS, La Salle, Universitat Ramon Llull, Barcelona, Spain\\
% ...
}
\end{flushleft}
%%%%%%%%%%%%%%%%%%%%%%%%%%%%%%%%%%%%%%%%%%

%% file: introduction.tex
Hadrons are systems bound by the strong interaction,
described at the fundamental level by quantum chromodynamics (QCD). 
While QCD is well understood at high energy in the 
perturbative regime, low-energy phenomena such as the binding of 
quarks and gluons within hadrons are more difficult to predict.
Several models and techniques, such as constituent quark models
or lattice QCD calculations, attempt to reproduce the spectrum of
the measured hadron masses (for a review, see Ref.~\cite{QuarkModel_PDG2012}).
While the masses of all expected ground-state mesons are now well measured, 
baryon data are still sparse. 
In particular, only six out of the sixteen 
$b$-baryon ground states predicted by the quark model have 
been observed so far~\cite{PDG2012}.
A complete and reliable experimental mass spectrum 
would allow for precision tests of a variety of QCD models~%
\cite{Karliner:2009,*Day:2012yh,*Liu:2007fg,*Jenkins:2007dm,*Roncaglia:1995az,*Mathur:2002ce,*Ebert:2005xj}.

The mass measurement of the heaviest observed $b$ baryon, the \myOb state
with $bss$ valence quark content, is of particular interest. 
While both the \dzero and \cdf collaborations have claimed
the observation of the $\myOb \to \jpsi \myO$ decay, 
the reported mass values, 
% $M(\myOb) = 
${\rm 6165   \pm 10 \, (stat) \pm 13 \, (syst)} \mevcc$
from \dzero~\cite{Abazov:2008qm} and 
% $M(\myOb) = 
${\rm 6054.4 \pm  6.8 \, (stat) \pm 0.9 \, (syst)} \mevcc$
from \cdf~\cite{Aaltonen:2009ny}, 
differ by more than 6 standard deviations.
On the other hand, there is good agreement between the mass measurements 
of the \myXb ($bsd$) baryon, which has also been observed by \dzero~\cite{Abazov:2007am}
and \cdf~\cite{Aaltonen:2007ap} in the $\myXb\to\jpsi\myX$ mode
and, more recently, by CDF~\cite{Aaltonen:2011wd}
in the $\myXb\to\myXcz\pim$ mode.
These measurements average to $5791.1\, \pm \, 2.2\mevcc$~\cite{PDG2012}.

This letter presents mass measurements of the weakly decaying \myLb ($bud$), \myXb and \myOb baryons
using the decay modes $\myLb\to\jpsi\myL$, $\myXb\to\jpsi\myX$ and $\myOb\to\jpsi\myO$
(charge-conjugated modes are implied throughout).
The mass differences with respect to the \myLb mass are also reported.
This analysis uses data corresponding to an integrated luminosity of $\rm 1.0~fb^{-1}$ and
collected in $pp$ collisions at a centre-of-mass energy of
$\sqrt{s}=7$~TeV with the \lhcb detector in 2011.

The \lhcb detector~\cite{LHCbDetector} is a single-arm forward
spectrometer covering the \mbox{pseudorapidity} range $2<\eta <5$,
designed for the study of particles containing \bquark or \cquark
quarks. The detector includes a high precision tracking system
consisting of a silicon-strip vertex detector surrounding the $pp$
interaction region, a large-area silicon-strip detector located
upstream of a dipole magnet with a bending power of about
$4{\rm\,Tm}$, and three stations of silicon-strip detectors and straw
drift tubes placed downstream. The combined tracking system has a
momentum resolution $\Delta p/p$ that varies from 0.4\% at 5\gevc to
0.6\% at 100\gevc, and an impact parameter resolution of 20\mum for
tracks with high transverse momentum. Charged hadrons are identified
using two ring-imaging Cherenkov detectors. Photon, electron and
hadron candidates are identified by a calorimeter system consisting of
scintillating-pad and preshower detectors, an electromagnetic
calorimeter and a hadronic calorimeter. Muons are identified by a
system composed of alternating layers of iron and multiwire
proportional chambers. The trigger~\cite{Aaij:2012me} consists of a
hardware stage, based on information from the calorimeter and muon
systems, followed by a software stage which applies a full event
reconstruction.

%% file: scale.tex
\begin{figure}[t]
\centering
\ifthenelse{\equal{\TwoColumns}{True}}{
\includegraphics[width=\columnwidth] {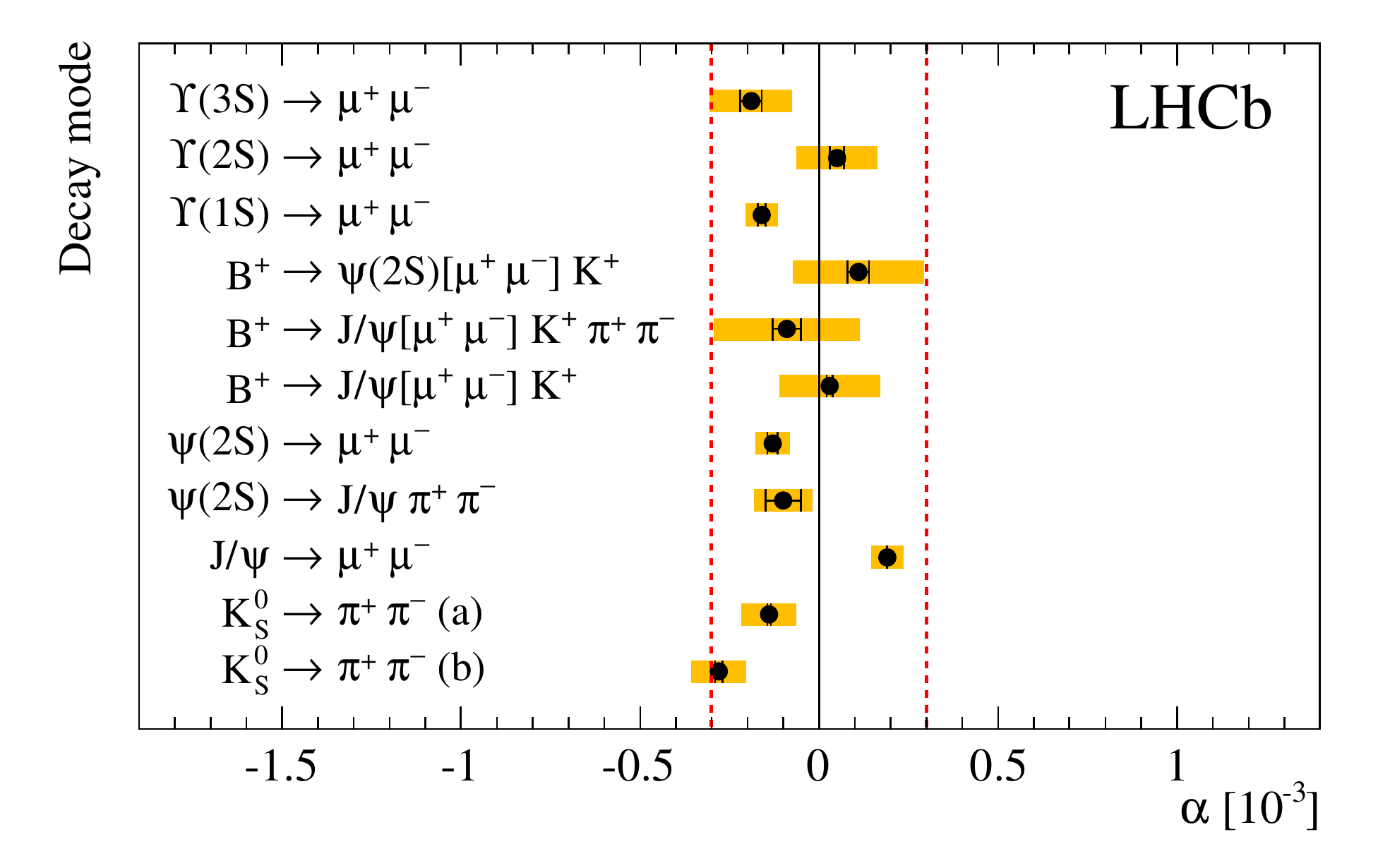}
}{
\includegraphics[width=0.75\columnwidth] {alpha17New.pdf}
}
\caption{\small Average momentum scale bias $\alpha$ determined from the 
reconstructed mass of various 
decay modes after the momentum calibration procedure.
The $\KS$ decays are divided into two categories according to whether both
daughter tracks (a) have hits or (b) do not have hits in the vertex detector.
The black error bars represent the statistical
uncertainty whilst the (yellow) filled areas also include contributions 
to the systematic uncertainty from the fitting procedure, the effect of 
QED radiative corrections, and the uncertainty
on the mass of the decaying meson~\cite{PDG2012}.
The (red) dashed lines show the assigned uncertainty of $\pm 0.3 \times 10^{-3}$ 
on the momentum scale.
}
\label{fig:scale}
\end{figure}

Precision mass measurements require the momenta of the final state particles to be determined accurately.
Therefore, an important feature of this analysis is the calibration of the tracker response.
This accounts for imperfect knowledge of the magnetic field and tracker alignment~\cite{LHCb-PAPER-2011-035}.
In order to reduce these dominant contributions to the systematic uncertainty,
a two-step momentum calibration procedure is applied.
Firstly, inclusive $\jpsi\to\mup\mun$ decays are used to account for
the changes in the relative momentum scale between different data taking periods.
Secondly, the absolute scale 
is derived from $B^+ \to \jpsi K^+$ decays,
taking the known masses~\cite{PDG2012} as references.
In this procedure, the use of
a \jpsi mass constraint allows the momentum scale to be 
determined as a function of the $K^+$ track kinematics. 
The resulting calibration is checked
with a variety of fully reconstructed decays listed in Fig.~\ref{fig:scale}. 
For each mode the mass distribution is modelled taking into account the effect
of QED radiative corrections, resolution and background, and the mean mass value is determined.
Following the procedure described in Ref.~\cite{LHCb-PAPER-2011-035}, the deviation of the measured
mass from the expected value is converted into an
estimate of an average momentum scale bias independent
of time and track kinematics. The bias is referred to as $\alpha$, which is defined such that
the measured mass becomes equal to the expected value
if all particle momenta are multiplied by $1-\alpha$.
By definition, one expects $\alpha=0$ to a good precision
for the $B^+ \to \jpsi K^+$ calibration mode.
Figure~\ref{fig:scale} shows the resulting values of $\alpha$. % for the various decay modes studied.
The nominal mass measurements are performed with $\alpha=0$ and
the largest value of $|\alpha|$ amongst the considered modes, $0.3 \times 10^{-3}$,
is conservatively taken as the systematic uncertainty on the calibrated momentum scale.
This uncertainty is somewhat larger than the $0.2 \times 10^{-3}$ achieved with the 2010 data~\cite{LHCb-PAPER-2011-035}, due to changes in the alignment of the tracking devices.

%% file: alt-selection.tex
In this analysis, the $b$-baryon decays $\myLb\to\jpsi\myL$, $\myXb\to\jpsi\myX$ and $\myOb\to\jpsi\myO$,
followed by $\jpsi\to\mup\mun$, $\myL\to\proton\pim$, $\myX\to\myL\pim$ and $\myO\to\myL\Km$, are reconstructed.
The topology of these decays is characterised by the long-lived particles in the decay chain.
The lifetime of weakly-decaying $b$ baryons is $\sim$1.5~ps and
their decay vertex is expected to be separated from the primary
$pp$ interaction vertex by $\sim$6~mm on average.
The \jpsi candidates are reconstructed from pairs of oppositely-charged tracks, originating
from the secondary vertex, that have hits in the muon detector.
In the \myXb (\myOb) decay chain the long-lived \myX (\myO) decays into a
\myL and a charged pion (kaon) at a tertiary vertex and the \myL decays at a quaternary vertex.
Since $\sim$90$\%$ of the decays are not fully contained in the vertex detector,
tracks that have no hits in the vertex detector are also considered in the reconstruction of the
tertiary and quaternary vertices. 

The selections of \myXb and \myOb candidates
are identical apart from the choice of the \myX, \myXb, \myO, \myOb invariant
mass ranges and particle identification requirements on the pion (kaon) from
the \myX (\myO) vertex. The \myLb selection is slightly different, owing to 
the different topology.

The \jpsi candidates are required to satisfy $|M_{\mu\mu}-M_{J/\psi}|<4.2\sigma$
where $M_{\mu\mu}$ is the reconstructed di-muon mass, $M_{J/\psi}$ the
$J/\psi$ mass~\cite{PDG2012} and $\sigma$ the estimated event-by-event uncertainty on $M_{\mu\mu}$ (typically $10\mevcc$).
The invariant mass windows for the \myL, \myX and \myO candidates are
$\pm 6\mevcc$, $\pm 11\mevcc$ and $\pm 11\mevcc$ around the expected masses~\cite{PDG2012}, respectively.
Particle identification requirements are applied to the kaon from the \myO candidate and the proton from the
\myL decay to improve the purity of the selected daughter particles, but none is placed on the pion from the \myX candidate.
In addition, the hyperon decay vertices are required to be downstream of the $b$-hadron decay vertex.

The \myLb, \myXb and \myOb mass resolutions are improved by performing a fit of
the decay topology and vertices~\cite{Hulsbergen} while constraining the masses
of the \jpsi, \myL, \myX and \myO hadrons to have their known values~\cite{PDG2012}, the
final-state and intermediate long-lived particles to originate from common vertices according to the decay chain,
and the $b$ baryon to originate from the primary vertex.
Three additional variables are considered for the selection.
These are the $\chi^2$ per degree of freedom ($\chi^2/{\rm ndf}$) from the fit,
the reconstructed decay time and the $\chi^2_{\rm IP}$ of the $b$ baryon from the primary vertex.
The $\chi^2_{\rm IP}$ is defined as the difference in the $\chi^2$ of the primary vertex fit with and without the $b$-baryon candidate.
In the case of the \myXb candidates, the selection requirements for these variables are chosen to maximise the expected significance of the \myXb signal;
the same selection is used for the \myOb candidates.
To determine the significance for a set of selection criteria the background yield is estimated from the yield of \myXb candidates found
in mass side-bands in the ranges 5600--5700\mevcc and 5900--6100\mevcc.
The expected signal yield is estimated using the product of the world average hadronisation fraction for $b \to \myXb$
and branching fractions for $\myXb \to \jpsi \myX$ and subsequent daughter particle decays~\cite{PDG2012},
the $b\bar{b}$ production cross-section in the \lhcb acceptance~\cite{LHCb-PAPER-2010-002} and the
selection efficiencies obtained from simulation.
The selection criteria giving the highest expected signal significance
correspond to a decay time greater than 0.25~ps, a $\chi^2/{\rm ndf}$ smaller than 4 and a $\chi^2_{\rm IP}$ smaller than 16.
Amongst these, the decay time requirement is the most powerful given the high level of background close to the interaction point.
In the case of the \myLb candidates, the decay time is required to be greater than 0.3~ps and the $\chi^2/{\rm ndf}$ smaller than 5
(no requirement on the $\chi^2_{\rm IP}$ is made).
The possibility of a cross-feed background between
\myXb and \myOb is investigated using simulation and found
to be negligible in comparison with the combinatorial background.

%% file: fits.tex
\begin{figure}[t]
\centering
\ifthenelse{\equal{\TwoColumns}{True}}{
\includegraphics[width=\columnwidth] {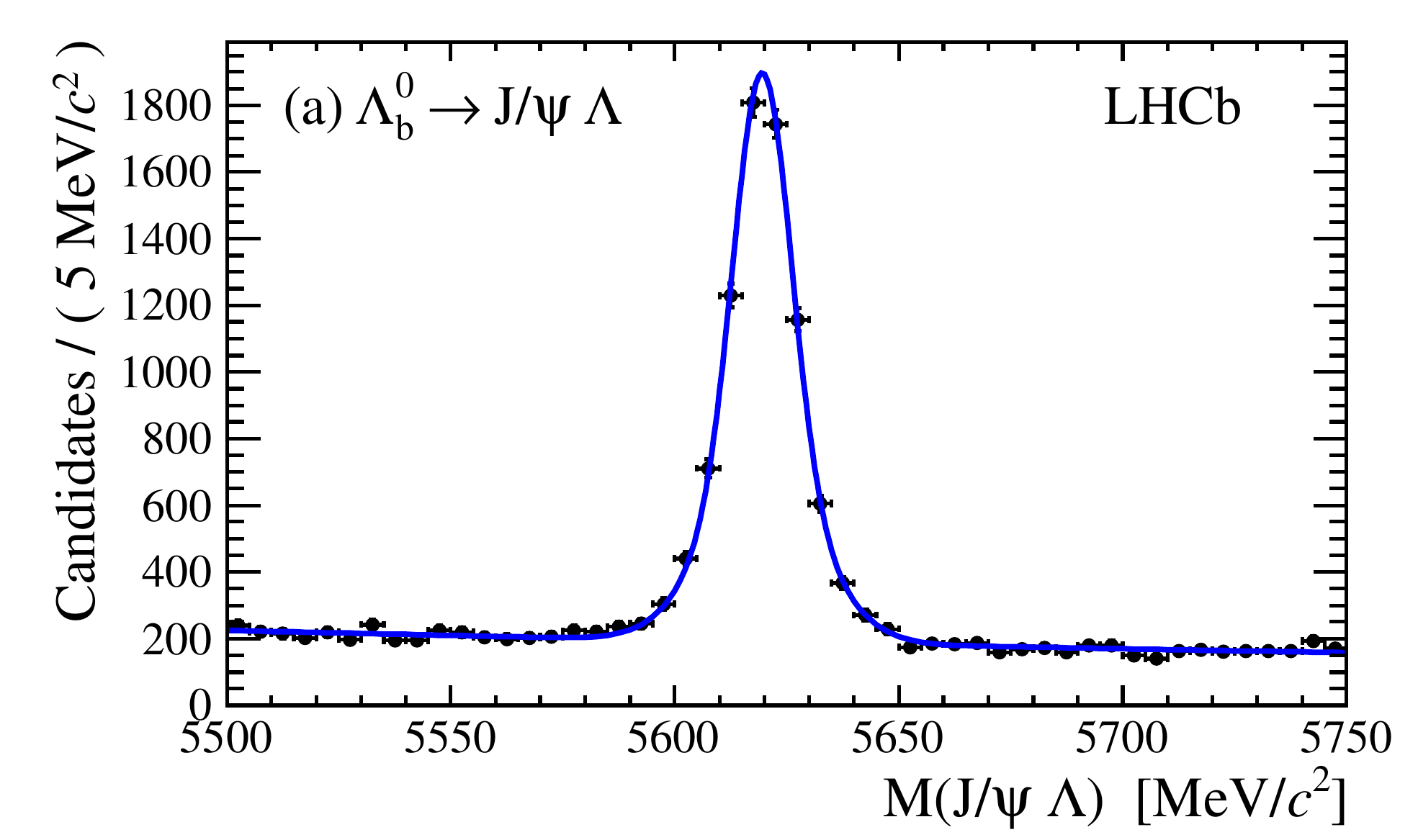}\\
\includegraphics[width=\columnwidth] {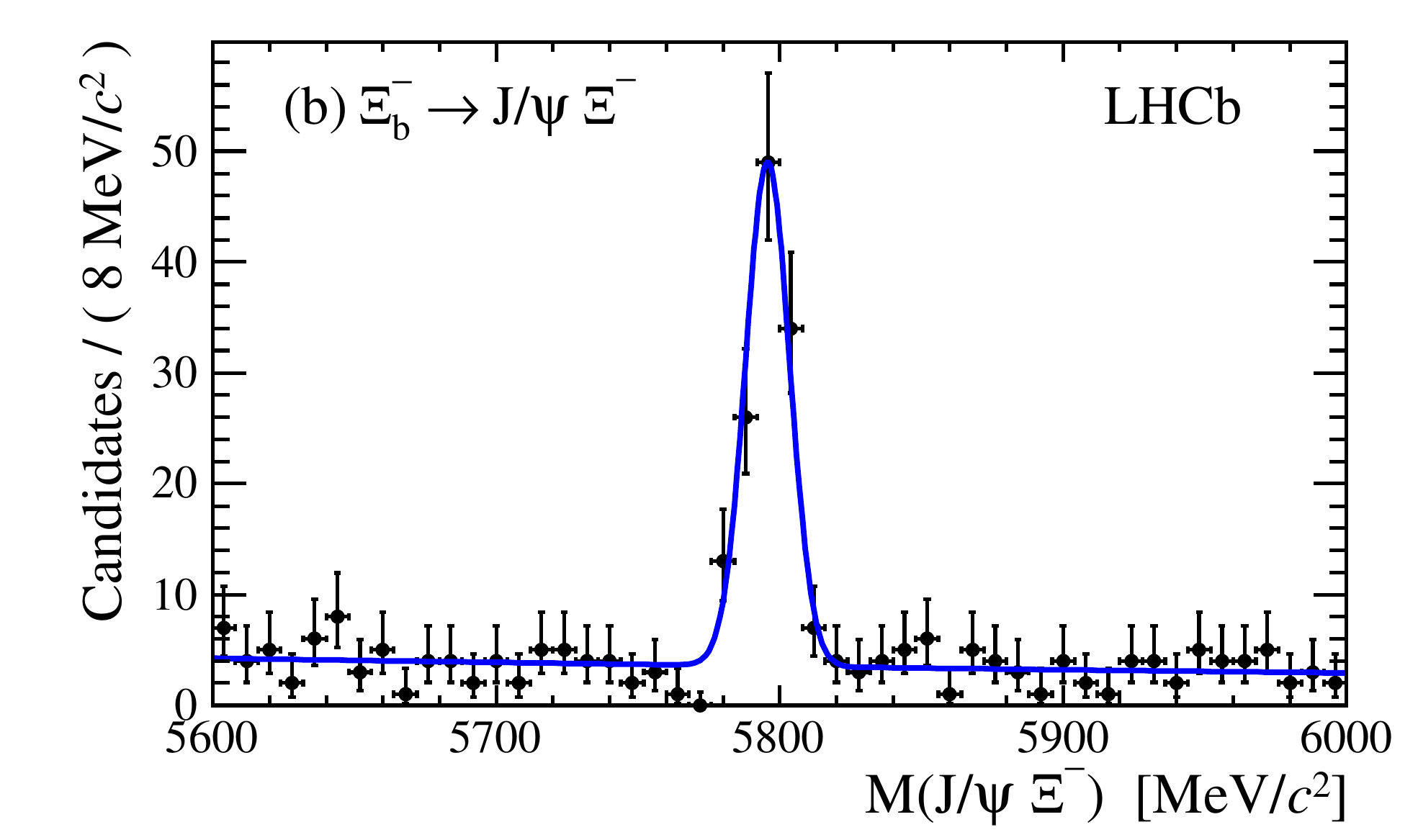}\\
\includegraphics[width=\columnwidth] {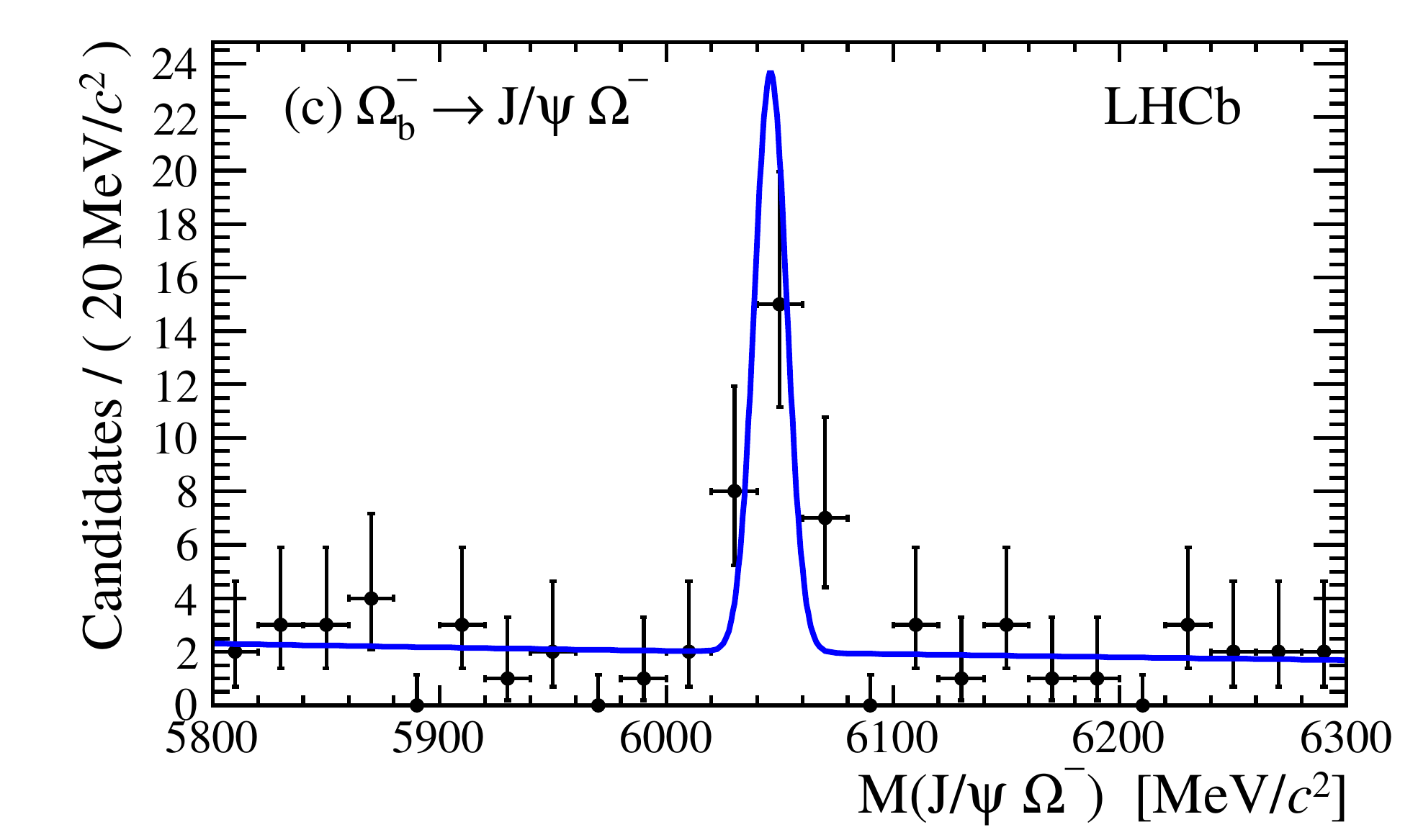}
}{
\includegraphics[width=0.5\columnwidth] {Lambda_b_Mass_Fit.pdf}%
\includegraphics[width=0.5\columnwidth] {Xi_b_Mass_Fit.pdf}\\
\includegraphics[width=0.5\columnwidth] {Omega_b_Mass_Fit.pdf}
}
\caption{\small Invariant mass distribution for (a) $\myLb \to \jpsi \myL$, (b) $\myXb \to \jpsi \myX$ and (c) $\myOb\to\jpsi\myO$
candidates. The results of the unbinned maximum likelihood fits are shown with solid lines.}
\label{fig:fits}
\end{figure}

The invariant mass distributions of the selected \myLb, \myXb and \myOb candidates
are shown in Fig.~\ref{fig:fits}.
In each case, the mass is 
measured by performing an unbinned extended maximum likelihood fit.
The \myLb, \myXb, \myOb candidates are retained for the mass fit if their invariant mass
lies in the range 5500--5750, 5600--6000, 5800--6300 \!\mevcc, respectively.
The signal component is described with a single Gaussian function
(or the sum of two Gaussian functions with common mean in the case of the \myLb baryon) 
and the background is modelled with an exponential function.
The widths of the \myLb and \myXb signals are left unconstrained in the fit.
Due to the low expected yield for the \myOb signal, the width
of the Gaussian function describing the \myOb signal is fixed
to the measured \myXb signal width multiplied 
by the ratio of \myOb and \myXb widths from the simulation
(8.2~\!\mevcc for \myOb and 8.9~\!\mevcc for \myXb).
The fit results are given in Table~\ref{tab:yields}.

\begin{table}[t]
\caption{\small Results of the fits to the invariant mass distributions.
The quoted uncertainties are statistical.
The \myLb signal is described by a double Gaussian function with widths $\sigma_1$ and $\sigma_2$;
the fraction of the yield described by the first component is $0.58 \pm 0.11$.}
\label{tab:yields}
\vspace{1.5ex}
\begin{center}
\begin{tabular}{l|c|c|c}
                       & Signal yield                    & Mass [\!\mevcc\!]                    & Width(s) [\!\mevcc\!]     \\
\hline
\multirow{2}{*}{\myLb} & \multirow{2}{*}{$6870 \pm 110$} & \multirow{2}{*}{$5619.53 \pm 0.13$}  & $\sigma_1 = ~\,6.4 \pm 0.5$ \\
                       &                                 &                                      & $\sigma_2 = 12.5 \pm 1.3$ \\
\myXb                  & $111  \pm 12$                   & $5795.8 \pm 0.9 $                    & $7.8 \pm 0.7$             \\
\myOb                  & $19   \pm 5 $                   & $6046.0 \pm 2.2 $                    & $7.2$ (fixed)             \\
\end{tabular}
\end{center}
\end{table}

The statistical significance of the \myOb signal is determined using simulated 
pseudo-experiments with background only.
We determine the probability that, anywhere in the mass range between 5800 and 6300~\!\mevcc, 
a peak appears with the expected width and a yield at least as large as that observed in the data.
This probability corresponds to 6 standard deviations, which we interpret as the statistical significance 
of the \myOb signal.

%% file: systematics.tex
The systematic uncertainties are evaluated by repeating the complete analysis
(including the track fit and the momentum scale calibration
when needed), varying in turn within its uncertainty 
each parameter to which the mass determination is sensitive.
The observed changes in the central values of the fitted masses relative
to the nominal results are then assigned as systematic uncertainties
and summed in quadrature. The systematic uncertainties are summarised in
Table~\ref{tab:MassSyst}.

The dominant systematic uncertainty is due to the momentum scale calibration described previously,
which is assigned an uncertainty of $\pm 0.3 \times 10^{-3}$.
A significant contribution to this uncertainty comes from the overall detector
length scale along the beam axis, which is known to a relative precision of $10^{-3}$~\cite{LHCb-PAPER-2011-010}.
This translates into a $\pm 0.13 \times 10^{-3}$ uncertainty on the momentum
scale, and is included in the overall $\pm 0.3 \times 10^{-3}$ uncertainty.
Most of the uncertainty related to the momentum scale is
removed in the measurements of the mass differences.

The uncertainty on
the amount of material assumed
in the track reconstruction for the energy loss ($dE/dx$) correction has been found to be small~\cite{LHCb-PAPER-2011-035}.
It translates into an uncertainty on the \myLb mass of 0.09\mevcc, which we apply to all masses.

The invariant mass of \myOb candidates is computed assuming the 
central value of the \myO world-average mass~\cite{PDG2012}. The uncertainty 
of $\pm 0.29\mevcc$ on this value is propagated as a systematic
uncertainty. A similar, but smaller, uncertainty is estimated for the \myXb and \myLb masses
from the imperfect knowledge of the \myX and \myL masses, respectively.  

Two alternative fits for the \myLb signal are performed: a first fit
where the candidates are split into two categories depending on
whether the daughter tracks have vertex detector information or not,
each category being described with a single Gaussian function
where the two Gaussian functions have a common mean, and a second fit using the
sum of two Crystal Ball functions~\cite{Skwarnicki:1986xj} with common peak value and otherwise unconstrained parameters.
The second fit allows to take into account possible QED radiative corrections.

The \myXb mass fit is repeated using as an alternative model either the sum
of two Gaussian functions with a common mean, or 
a single Crystal Ball function. In the \myOb mass fit, the
fixed Gaussian width is varied within both the uncertainty of the
fitted \myXb width and the statistical uncertainty
of the width ratio from simulation.

An alternative background model assuming a linear
shape leads to negligible changes.
We also repeat the \myXb and \myOb mass fits in a restricted mass range of 
5650--5950\mevcc and 5900--6200\mevcc, respectively, 
and assign the resulting change as a systematic uncertainty.

\begin{table}[t]
\caption{\small Systematic uncertainties (in \!\mevcc) on the 
mass measurements and their differences. The total systematic uncertainty is obtained from adding all uncertainties in quadrature.}
\label{tab:MassSyst}
\vspace{1.5ex}
\begin{center}
\begin{tabular}{l|c|c|c|c|c}
Source & $\myLb$ & $\myXb$ & $\myOb$ & $\myXb$--$\myLb$ & $\myOb$--$\myLb$ \\
\hline
Momentum scale                          & 0.43  & 0.43 & 0.31 & 0.01 & 0.12 \\
$dE/dx$ correction                      & 0.09  & 0.09 & 0.09 & 0.01 & 0.01 \\
Hyperon mass                            & 0.01  & 0.07 & 0.25 & 0.07 & 0.25 \\
Signal model                            & 0.07  & 0.01 & 0.24 & 0.07 & 0.25 \\
Background model                        & 0.01  & 0.01 & 0.02 & 0.01 & 0.02 \\
\hline
Total                                   & 0.45  & 0.45 & 0.47 & 0.10 & 0.37 \\
\end{tabular}
\end{center}
\end{table}

%% file: conclusion.tex
In summary, the \myLb, \myXb and \myOb baryons are observed in the $\myLb\to\jpsi\myL$, $\myXb\to\jpsi\myX$ and 
$\myOb\to\jpsi\myO$ decay modes using 1.0~fb$^{-1}$ of
$pp$ collisions collected in 2011 at a centre-of-mass energy of $\sqrt{s}=7$~TeV.
The statistical significance of the observed $\myOb\to\jpsi\myO$ signal is 6 standard deviations. 
The masses of the $b$ baryons are measured to be
\begin{eqnarray*}
M(\myLb)\, & = & 5619.53   \pm 0.13   \pm 0.45   \mevcc  , \\
M(\myXb)   & = & 5795.8~\, \pm 0.9~\, \pm 0.4~\, \mevcc  , \\
M(\myOb)   & = & 6046.0~\, \pm 2.2~\, \pm 0.5~\, \mevcc  ,
\end{eqnarray*}
where the first (second) quoted uncertainty is statistical (systematic).
The dominant systematic uncertainty, due to the knowledge of the momentum scale, partially 
cancels in mass differences. We obtain
\begin{eqnarray*}
M(\myXb)-M(\myLb) & = & 176.2 \pm 0.9 \pm 0.1 \mevcc, \\
M(\myOb)-M(\myLb) & = & 426.4 \pm 2.2 \pm 0.4 \mevcc.
\end{eqnarray*}

A measurement of the \myLb mass based on the 2010 data sample, 
$M(\myLb) = \rm 5619.19 \pm 0.70 \pm 0.30$\mevcc,
has been previously reported by LHCb~\cite{LHCb-PAPER-2011-035}.
Since the new alignment and momentum calibration procedures differ from
those applied in the previous study, a possible correlation between the 
systematic uncertainties related to the momentum scale can be neglected.
Considering that the only correlated systematic uncertainties are those due to energy loss correction and mass fitting,
the weighted average of the two \myLb mass measurements that minimizes the total uncertainty is
$$ M(\myLb) = 5619.44 \pm 0.13 \pm 0.38 \mevcc \,.$$ % = 5619.44 \pm0.40\, (tot)\mevcc \,.$$

These \myLb, \myXb and \myOb mass measurements are the most precise to date.
They are compared in Table~\ref{tab:average} with the single most precise measurements
from ATLAS, CDF and \dzero, and with the current world averages~\cite{PDG2012}.
The \myLb and \myXb results are in agreement with previous measurements.
The \myOb result is in agreement with the \cdf measurement~\cite{Aaltonen:2009ny},
but in disagreement with the \dzero measurement~\cite{Abazov:2008qm}.

\begin{table}
\caption{\small Comparison of the $b$-baryon mass measurements using the full 2011 data sample with
the single most precise results from the ATLAS~\cite{Aad:2012shb}, CDF~\cite{Aaltonen:2009ny,Acosta:2005mq}
and \dzero~\cite{Abazov:2008qm,Abazov:2007am} collaborations, and with the PDG averages~\cite{PDG2012}.
The PDG averages contain the results from CDF and \dzero as well as the \myLb measurement from LHCb performed with the 2010 data sample.
The quoted errors include statistical and systematic uncertainties. All values are in\,\mevcc.}
\vspace{1.5ex}
\centering
\begin{tabular}{l|l|l|l}
            &  \multicolumn{1}{c|}{$M(\myLb)$}
            &  \multicolumn{1}{c|}{$M(\myXb)$}
            &  \multicolumn{1}{c }{$M(\myOb)$} \\ 
       \hline
      ATLAS &  $5619.7 \pm 1.3~$ & --                & --                \\
       CDF  &  $5619.7 \pm 1.7~$ & $5790.9  \pm 2.7$ & $6054.4  \pm 6.9$ \\ 
    \dzero  &  --                & $5774~~\,\pm 19 $ & $6165~~\,\pm 16$  \\ 
       PDG  &  $5619.4 \pm 0.7~$ & $5791.1  \pm 2.2$ & $6071~~\,\pm 40$  \\ 
\hline
       LHCb &  $5619.5 \pm 0.5~$ & $5795.8  \pm 1.0$ & $6046.0  \pm 2.3$ \\ 
\end{tabular}
\label{tab:average}
\end{table}

%% file: acknowledgements.tex
\section*{Acknowledgements}

\noindent We express our gratitude to our colleagues in the CERN
accelerator departments for the excellent performance of the LHC. We
thank the technical and administrative staff at the LHCb
institutes. We acknowledge support from CERN and from the national
agencies: CAPES, CNPq, FAPERJ and FINEP (Brazil); NSFC (China);
CNRS/IN2P3 and Region Auvergne (France); BMBF, DFG, HGF and MPG
(Germany); SFI (Ireland); INFN (Italy); FOM and NWO (The Netherlands);
SCSR (Poland); ANCS/IFA (Romania); MinES, Rosatom, RFBR and NRC
``Kurchatov Institute'' (Russia); MinECo, XuntaGal and GENCAT (Spain);
SNSF and SER (Switzerland); NAS Ukraine (Ukraine); STFC (United
Kingdom); NSF (USA). We also acknowledge the support received from the
ERC under FP7. The Tier1 computing centres are supported by IN2P3
(France), KIT and BMBF (Germany), INFN (Italy), NWO and SURF (The
Netherlands), PIC (Spain), GridPP (United Kingdom). We are thankful
for the computing resources put at our disposal by Yandex LLC
(Russia), as well as to the communities behind the multiple open
source software packages that we depend on.